\documentclass[onecolumn,amsmath,amssymb,nofootinbib,12pt]{article}
\usepackage{jheppub}
\usepackage{ifpdf}
\usepackage{graphicx,subcaption}
\usepackage{amsfonts}
\usepackage{amsmath}
\usepackage{amssymb}
\usepackage{epsfig}
\usepackage{pdfpages}
\usepackage{graphicx,epstopdf}
\usepackage[makeroom]{cancel}
\hypersetup{pdftex,colorlinks=true,allcolors=blue}
\usepackage{array}
\usepackage{ulem}

\newcommand{\fig}[1]{Fig.\ref{#1}}
\newcommand{\bra}[1]{|#1\rangle}
\def\be{\begin{equation}}
\def\ee{\end{equation}}
\def\ba{\begin{eqnarray}}
\def\ea{\end{eqnarray}}

\def\nn{\nonumber}
\def\lf{\left}
\def\rt{\right}

\newcommand{\eq}[1]{(\ref{#1})}

\makeatletter

\newcommand{\Rmnum}[1]{\expandafter\@slowromancap\romannumeral #1@}
\makeatother
\def\nn{\nonumber}\def\lf{\left}\def\rt{\right}\def\q{\theta} \def\w{\omega}  \def\y {\psi}   \def\p {\pi} \def\a {\alpha} \def\s {\sigma} \def\d {\delta} \def\f {\phi}  \def\h {\eta}   \def\l {\lambda}       \def\pd {\partial} \def \inf {\infty}  
\def\Q{\Theta} \def\W{\Omega} \def\Y {\Psi}    \def\S {\Sigma} \def\D {\Delta} \def\F {\Phi}  \def\L {\Lambda}    \def\grad{\nabla}\def\.{\cdot}
\def\math {\mathcal}
\usepackage{xspace}
\begin{document}
\title{\Large Modified ``complexity equals action" conjecture}
\author[a]{Jie Jiang, }
\affiliation[a]{Department of Physics, Beijing Normal University,
Beijing 100875, China}
\author[b]{Xiao-Wei Li}
\affiliation[b]{School of Physics, University of Chinese Academy of Sciences, Beijing 100049, China}
\emailAdd{jiejiang@mail.bnu.edu.cn, lixiaowei16@mails.ucas.ac.cn}
\date{\today}
\abstract{
In this paper, we first use the ``complexity equals action"  conjecture to discuss the complexity growth rate in both perturbation Einsteinian cubic gravity and non-perturbation Einstein-Weyl gravity. We find that the CA complexity rate in these cases is divergent. To avoid this divergence, we modify the original conjecture, where we assume that the complexity of the boundary state equals the boundary actions contributed by the null segments as well as the joints of the Wheeler-DeWitt patch. Then, the late time growth rate of this modified holographic complexity is given by entropy $S$ times temperature $T$, which is quite in agreement with the circuit analysis. Finally, to test its rationality, we also investigate the switchback effect by evaluating it in a Vaidya geometry and analyze the results in circuit models.
}
\maketitle
\section{Introduction}\label{se1}
Recently, applications of quantum information concepts to high energy physics and gravity have led to many promising results. Among all, it has become more clear that special properties of entanglement in holographic quantum field theories are account for the emergence of smooth higher-dimensional geometries in gauge-gravity duality\cite{Maldacena:2013xja,Aharony:1999ti,Takayanagi:2016,VanRaamsdonk:2010pw}. This faith inspired a further deeper idea that ``ER=EPR''\cite{Maldacena:2013xja} when consider a  non-traversable wormhole created by an Einstein-Rosen (ER)\cite{Einstein:1935} bridge and a pair of maximally entangled black holes, based on the point of view that a black hole might be highly entangled with a system that is effectively infinitely far away. EPR denotes the quantum entanglement (Einstein-Podolsky-Rosen paradox). However, recent developments pointed out that holographic entanglement is unable to describe the bulk spacetime far behind the event horizon of black holes \cite{Susskind:2014moa}. Hence there must exist some other quantity we yet do not know which have the information about internal of black hole. And it has been suggested by Susskind and collaborators that this quantity is the quantum computational complexity of the boundary state, which presents the minimum number of elementary operations or gates from a reference state to a target state.

There are two related proposals conjectured to capture the complexity of the boundary state in holography: ``complexity equals volume" (CV) \cite{L.Susskind,D.Stanford} and ``complexity equals action" (CA) \cite{BrL,BrD}. These conjectures have attracted many researchers to investigate the properties of both holographic complexity and circuit complexity in quantum field theory, $e.g.$,\cite{A37,A38,A39,A1,A2,A3,A4,A5,A6,A7,A8,A9,A10,A11,A12,A13,A14,A15,A16,A17,A18,A19,A20,A21,A22,A23,A24,A25,A26,A27,A28,A29,A30,A31,A32,A33,A34,A35,A36,A51,Fan:2019mbp}
In this paper, we only focus on the CA conjecture, which states that the complexity of a particular state $|\y(t_L,t_R)\rangle$ on the AdS boundary is given by
\ba
\math{C}\lf(|\y(t_L,t_R)\rangle\rt)\equiv\frac{I_\text{WDW}}{\p\hbar}\,,
\ea
where $I_\text{WDW}$ is the on-shell action in the corresponding Wheeler-DeWitt (WDW) patch, which is enclosed by the past and future light sheets sent into the bulk spacetime from the timeslices $t_L$ and $t_R$. As argued in \cite{Lloyd}, there is a bound of the complexity growth rate of a system of energy $E$ at the late time
\ba
\dot{\math{C}}\leq \frac{2 E}{\p \hbar}\,,
\ea
which may be thought of as the Lloyd's bound on the quantum complexity. Although some recent developments have cast the exact prefactor of this bound into question, it is not difficult to believe that, for any system, the quantum complexity should always have a finite growth rate.

Recently, Nally \cite{Nally:2019rnw} investigated the CA conjecture in some stringy modes of higher curvature gravity where the general relativity is perturbed by higher powers of the Riemann tensor. Their results suggest that, once stringy effects are taken into account, the neutral black hole will have a divergent complexity growth rate under the first leading order correction. One of the reasonable interpretations mentioned by the author is that, in the CA picture, the complexity growth rate is a nonanalytic function of the small coupling parameter. Therefore, it is necessary to investigate whether the similar divergences would be found in the non-perturbation higher curvature theories. According to the calculation in next section, we find that the CA complexity rate is also divergent in the non-perturbtion Einstein-Weyl gravity. This apparent failure of the CA conjecture motivates us to re-define the holographic complexity such that the complexity growth rate is convergent when the higher curvature term is concerned.

Another motivation for the present paper is from the discussion of the quantum circuit model. By studying a simple class of systems known as random quantum circuits with $N$ qubits, it has been shown that for generic circuits, after a short period of transient initial behavior, the complexity grows linearly in time, and finally saturates at a maximum value. In the context of the AdS/CFT, we need set $N$ to be very large. Then, it can be generally argued that at the late times, this quantum complexity should continue to grow with a rate given by\cite{L.Susskind,D.Stanford}
\ba
\frac{d\math{C}}{dt}\sim TS\,,
\ea
where the entropy represents the width of the circuit and the temperature is an obvious choice for the local rate at which a particular qubit interacts. And in high temperature limit, this feature is held in the CA context. However, we also hope that this feature is held even at low temperature regimes.

The remainder of our paper is organized as follows: In Sec.\ref{se2}, we first review the results in \cite{Nally:2019rnw}, and then study the CA conjecture in both perturbed Einsteinian cubic gravity and non-perturbation Einstein-Weyl gravity. In Sec.\ref{se3}, in order to avoid the divergence mentioned in Sec.\ref{se2}, we modify the CA conjecture and investigate its time involution as well as the switchback effect for a general higher curvature gravitational theory. Finally, we conclude our paper in Sec.\ref{se4}.

\begin{figure}
\centering
\includegraphics[width=0.8\textwidth]{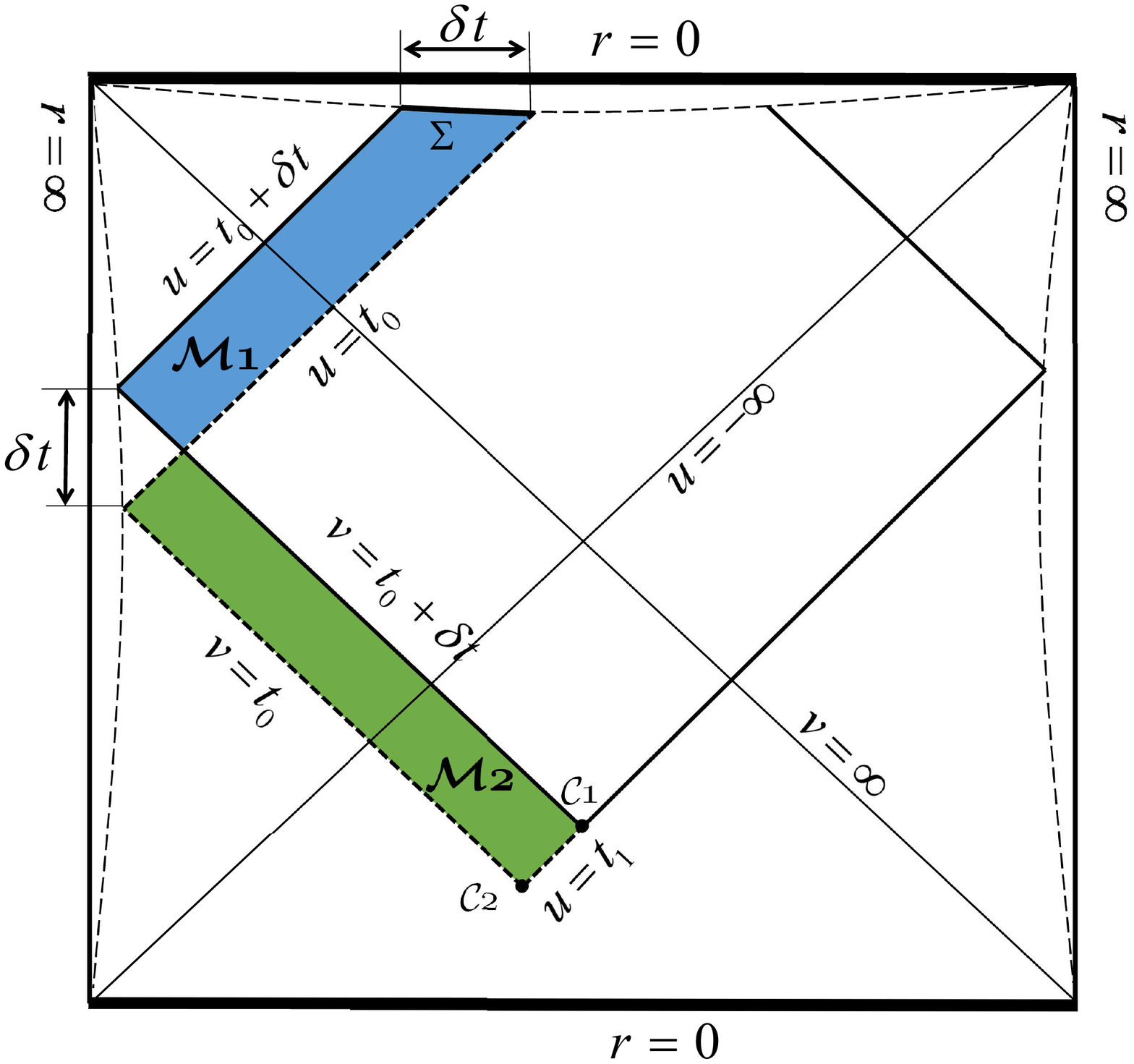}
\caption{ Wheeler-DeWitt patches of a Schwarzschild-AdS type black hole.}\label{WdW}
\end{figure}
\section{CA complexity for higher curvature gravity}\label{se2}
In \cite{Nally:2019rnw}, Nally studied the holographic complexity with CA conjecture for higher curvature gravitational theories which are directly motivated by string theory. In this case, string theory naturally comes equipped with an infinite tower of correction to the general relativity. And then, the corresponding bulk action can be expressed as
\ba\begin{aligned}
I=\int d^{d+2}x\sqrt{-g}\lf(R-2\L+\sum_{n=1}^{\inf}\a^n\math{L}_n\rt)\,,
\end{aligned}\ea
where $\math{L}_n$ is in general an $n$-th order polynomial in the Riemann tensor and possibly the gauge fields. After the calculation on a $5$-dimensional Gauss-Bonnet as well as Weyl gravitational theories, they found that the holographic complexity is extremely sensitive to singularity effects and the leading CA complexity growth rate is divergent at the late times. And this divergence comes from the singularity. However, the Lloyd bound suggests that any system, even if it has an infinite-dimensional Hilbert space, should have a finite complexity growth rate. Then, if the CA conjecture should prove correct in these perturbation gravitation theory, the complexity growth rate must be a nonanalytic function of the small parameter $\a$. This means that when all of the corrections is concerned, the CA complexity growth rate will be convergent. Therefore, it is also necessary to study whether these similar divergences would be found in a non-perturbation higher curvature theory. To better understand the divergences of the CA complexity growth rate, we next consider both the perturbation Einsteinian cubic gravity and non-perturbation Einstein-Weyl gravity.

\subsection{Perturbation Einsteinian cubic gravity}\label{se21}
In this subsection, we consider a $4$-dimensional  Einsteinian cubic gravity. As shown in \cite{BC,Jiang3}, its total action can be given by
\ba\begin{aligned}
&I=\int_\math{M}d^4x\sqrt{-g}\lf(R+\frac{6}{L^2} +\l \math{P}\rt)+4\sum_{s}\lf(\int_{\math{B}_s}d^3x\sqrt{|h|}\Y^{ab}K_{ab}\rt)\\
&+(-1)^\l \int_{\math{C}_{\l}}d^2x\sqrt{\s}\hat{\Y}\h_\l-\int_{\math{N}}d\l d^2x\sqrt{\s}\hat{\Q}\ln\lf(l_{\text{ct}}\Q\rt)d\l\,,
\end{aligned}\ea
with $l_{ct}$ an arbitrary length scale and
\ba\begin{aligned}
\math{P}=12 R_{acbd}R^{cedf}R_e{}^a{}_f{}^b+R_{ab}{}^{cd}R_{cd}{}^{ef}R_{ef}{}^{ab}-12R_{abcd}R^{ac}R^{bd}+8R_a{}^bR_b{}^cR^a{}_c\,.
\end{aligned}\ea
Here $\hat{\Q}=\nabla_a(k^a\hat{\Y})=\frac{1}{\sqrt{\s}}\pd_\l(\hat{\Y}\sqrt{\s})$ and $\Q=\nabla_ak^a=\frac{1}{\sqrt{\s}}\pd_\l(\sqrt{\s})$ is the expansion scalar of the null generator. And the auxiliary fields can be written as
\ba\begin{aligned}
\Y_{ab}=\y_{acbd}n^cn^d\,,\ \ \ \ \hat{\Y}=\y^{abcd}\epsilon_{ab}\epsilon_{cd}\,,
\end{aligned}\ea
where
\ba\begin{aligned}
\y_{abcd}&=\frac{\pd L}{\pd R^{abcd}}\\
&=\frac{1}{2}\lf(g_{ac}g_{bd}-g_{ad}g_{bc}\rt)+6\l \left(R_{ad}R_{bc}
-R_{ac}R_{bd}+g_{bd}R_a{}^eR_{ce}-g_{ad}R_b{}^eR_{ce}\right.\\
&-g_{bc}R_a{}^eR_{de}+g_{ac}R_b{}^eR_{de}-g_{bd}R^{ef}R_{aecf}+g_{bc}R^{ef}R_{aedf}+g_{ad}R^{ef}R_{becf}\\
&-g_{ac}R^{ef}R_{bedf}-3R_a{}^e{}_d{}^fR_{becf}
+3R_a{}^e{}_c{}^fR_{bedf}+\frac{1}{2}R_{ab}{}^{ef}R_{cdef})\,,
\end{aligned}\ea
$n^a$ is the outward-directed normal vector for the corresponding surface, and $\epsilon_{ab}$ is the binormal for the joints. Here this action includes not only the bulk action of the Einsteinian cubic gravity but the surface terms, corner terms and counter term as well. According to the bulk action, the equation of motion can be written as
\ba\begin{aligned}
\y_{acde}R_b{}^{cde}-\frac{1}{2}g_{ab}L-2\grad^c\grad^d\y_{acdb}=0\,,
\end{aligned}\ea
With similar consideration as \cite{Nally:2019rnw}, here we also treat $\l$ as a small parameter, i.e., the cubed terms are regarded as a small correction to Einstein gravity. Then, we can obtain the Schwarzschild-AdS-type black hole solution at the first subleading, whose line element can be shown as \cite{Mann}
\ba\label{ds2}
ds^2=-f(r)dt^2+\frac{dr^2}{f(r)}+r^2\lf(d\q^2+\sin^2\q d\f^2\rt)
\ea
with the blackening factor
\ba
f(r)=1-\frac{2M}{r}+\frac{r^2}{L^2}+\l\lf(-\frac{368M^3}{r^7}+\frac{216M^2}{r^6}+\frac{336M^2}{L^2r^4}+\frac{b}{2r}-\frac{8r^2}{L^6}\rt)\,.
\ea

Then, let us evaluate complexity growth rate with the CA conjecture. As illustrated in the Penrose diagram of the SAdS-type spacetime \fig{WdW}, $I(t_L,t_R)$, denoted as the action for the WDW patch determined by the time slice on the left and right AdS boundaries, is invariant under the time translation $I(t_L+\d t,t_R-\d t)=I(t_L,t_R)$. Thus the action growth can be computed as $\d I = I(t_0+\d t,t_1)-I(t_0,t_1)$, where the time on the right boundary has been fixed. To regulate the divergence near the AdS boundary, a cut-off surface $r=r_\L$ is introduced. In addition, we also introduce a spacelike surface $r=\epsilon$ to avoid running into the spacelike singularity inside of this neutral black hole. As such, we shall focus on the situation in which the boundary consists solely of null and spacelike segments only with spacelike joints. In addition, for simplicity we shall adopt the affine parameter for the null generator of null segments such that the surface term vanishes for null segments. With these in mind, we have
\ba\begin{aligned}\label{agr}
\d I &=I_{\d\math{M}}+I_{\S}+I_{\d\math{C}}+ \d I_{\text{ct}}\,.\\
\end{aligned}\ea
with $I_{\d\math{M}}=I_{\math{M}_1}-I_{\math{M}_2}$ and $I_{\d\math{C}}=I_{\math{C}_1}-I_{\math{C}_2}$.
Here $\math{M}_1$ is bounded by $u=t_0$, $u=t_0+\d t$,$v=t_0+\d t$, and $r=r_\text{min}$. $\math{M}_2$ is bounded by $u=t_0$, $v=t_0$, $v=v_0+\d t$, and $u=t_1$. The null coordinates are defined as $u=t+r^*(r)$ and $v=t-r^*(r)$ with
\ba
r^*(r)=-\int_r^\inf \frac{dr}{f(r)}\,.
\ea
With the above preparation, we first calculate the change of the action which is contributed by the bulk region. Through straight calculation, we have
\ba
I_{\d\math{M}}=\W_2\d t\left.\lf[-\frac{2r^3}{L^2}+\l\lf(-\frac{1952M^3}{r^6}+\frac{864M^2}{r^5}
+\frac{480M^2}{L^2r^4}+\frac{64r^3}{L^6}\rt)\rt]\right|_\epsilon^{r_1}\,,
\ea
where we denote $\W_2=4\p$. We can see that when the cut-off surface $r=\epsilon$ approach to the singularity, i.e., $\epsilon\to 0$, the bulk contributions will be divergent. Then, we proceed to the surface contribution from the spacelike surface $r=\epsilon$, we have
\ba\begin{aligned}
&I_\S=4\int_\math{S}d^3x\sqrt{h}\y^{acbd}n_cn_d\grad_a n_b\,,\\
&=\W_2\d t\lf[6M+\l\lf(-\frac{3b}{2}-\frac{144M}{L^4}-\frac{2400M^3}
{\epsilon^6}+\frac{960M^2}{\epsilon^5}-\frac{288M^2}{L^2\epsilon^3}+\frac{192M}{L^2\epsilon^2}\rt)\rt]\,,
\end{aligned}\ea
where $n^a=(dr)^a/\sqrt{|f|}$ is the outward-directed normal vector of $\math{S}$. And this term is also divergent.

To evaluate the corner contributions from $\math{C}_1$ and $\math{C}_2$, we first choose $k_{1a}=\grad_a u$ and $k_{2a}=-\grad_a v$ as the null generator of the past right and past left null boundaries separately. By using these expressions, we can obtain $k_1\cdot k_2=-2/f$ and
\ba
\hat{\Y}=-2+\l\lf(\frac{48}{L^4}-\frac{240M^2}{r^6}-\frac{192b}{L^2r^3}\rt)\,.
\ea
Then, we have
\ba\label{corner}\begin{aligned}
I_{\d\math{C}}&=\W_2\d t f'(r_1)\lf[r^2_1-\l\lf(\frac{24r_1^2}{L^2}-\frac{120M^2}{r_1^4}-\frac{96b}{L^2r_1}\rt)\rt]\\
&+\W_2\d t\lf[2r_1-\l\lf(\frac{48r_1}{L^2}+\frac{480M^2}{r_1^5}+\frac{96 b}{L^2r_1^2}\rt)\rt]f(r_1)\ln\lf(-f(r_1)\rt)\,,
\end{aligned}\ea
where we have used
\ba\label{dr}
\d r=r_1-r_2=-\frac{1}{2}f(r_1)\d t\,
\ea
with $r_2$ the $r$ coordinate of $\math{C}_2$.

Finally, we consider the conter term contributions. By the translation symmetry, there are only two null segments contributing to the action growth. The first one comes from the null segment $u=t_1$ with $r$ as the affine parameter, i.e., $k_1^{a}=\lf(\frac{\pd}{\pd r}\rt)^a$, which gives rise to the expansion $\Q=2/{r}$ and
\ba
\hat{\Q}=-\frac{4}{r}+\l\lf(\frac{960M^2}{r^7}+\frac{192M}{L^2r^4}+\frac{96}{L^4r}\rt).
\ea
Then, we can obtain
\ba\label{Sct1}\begin{aligned}
I_\text{ct}^{(1)}&=-\W_2\int_{r_1}^{r_\text{max}}drr^2\hat{\Q}\ln\left(l_\text{ct}\Q\right)\\
&=\text{``UV term"}+\W_2\lf[r_1^2-\l\lf(\frac{60M^2}{r^4_1}+\frac{192M}{L^2r_1}+\frac{24r_1^2}{L^4}\rt)\rt]\\
&+\W_2\lf[2r_1^2+\l\lf(\frac{240M^2}{r_1^4}+\frac{192M}{L^2r_1}-\frac{48r_1^2}{L^4}\rt)\rt]\ln\lf(\frac{2l_\text{ct}}{r_1}\rt)\,.
\end{aligned}
\ea
With similar calculation, we have $I_\text{ct}^{(2)}=I_\text{ct}^{(1)}$. Then, the counter term contributions is given by
\ba\label{dIct}
\d I_\text{ct}=2\W_2\d t\lf[2r_1-\l\lf(\frac{48r_1}{L^2}+\frac{480M^2}{r_1^5}+\frac{96 b}{L^2r_1^2}\rt)\rt]f(r_1)\ln\lf(\frac{2l_\text{ct}}{r_1}\rt)\,.
\ea
By summing all the previous results, one can obtain
\ba\begin{aligned}
\dot{I}&=\W_2\lf[6M-\frac{2r_1^3}{L^2}+r_1^2f'(r_1)+\lf[2r_1-\l\lf(\frac{48r_1}{L^2}+\frac{480M^2}{r_1^5}+\frac{96 b}{L^2r_1^2}\rt)\rt]f(r_1)\ln\lf(\frac{2l_\text{ct}}{r_1}\rt)\rt]\\
&+\W_2\l \lf[\frac{192M}{L^2\epsilon^2}-\frac{768M^2}{L^2\epsilon^3}+\frac{96M^2}{\epsilon^5}
-\frac{448M^3}{\epsilon^6}-\frac{144M}{L^4}-\frac{3b}{2}-\frac{1952M^3}{r_1^6}\rt]\\
&+\W_2\l\lf[\frac{864M^2}{r_1^5}+\frac{480M^2}{L^2r_1^4}+\frac{64r_1^3}{L^6}-f'(r_1)\lf(\frac{24r_1^2}{L^2}-\frac{120M^2}{r_1^4}-\frac{96b}{L^2r_1}\rt)\rt]\,.\\
\end{aligned}\ea
From this result, we can see that the CA complexity growth rate is divergent when the cut-off surface approaches singularity at the first leading order approximation of $\l$.

\subsection{Non-perturbation Einstein-Weyl gravity}\label{se22}
In this subsection, we investigate the CA conjecture for a non-perturbation $4$-dimensional Einstein-Weyl theory. The total action is given by\cite{Jiang3}
\ba\begin{aligned}
&I=\int_\math{M}d^4x\sqrt{-g}C_{abcd}C^{abcd}+4\sum_{s}\lf(\int_{\math{B}_s}d^3x\sqrt{|h|}\Y^{ab}K_{ab}\rt)\\
&+(-1)^\l \int_{\math{C}_{\l}}d^2x\sqrt{\s}\hat{\Y}\h_\l -\int_{\math{N}}d\l d^2x\sqrt{\s}\hat{\Q}\ln\lf(l_{\text{ct}}\Q\rt)\,,
\end{aligned}\ea
According to the bulk action, the equation of motion can be written as
\ba
(2\grad^b\grad^c+R^{bc})C_{abcd}=0\,.
\ea
One can verify that the conformal gravity admits a Schwarzschild-AdS black hole solution, whose line element can be shown as
\ba\label{ds2}
ds^2=-f(r)dt^2+\frac{dr^2}{f(r)}+r^2\lf(d\q^2+\sin^2\q d\f^2\rt)
\ea
with the blackening factor
\ba
f(r)=1-\frac{2M}{r}+\frac{r^2}{L^2}\,,
\ea
By using this line element \eq{ds2}, one can further obtain
\ba\begin{aligned}
C_{abcd}C^{abcd}=\frac{48M^2}{r^6}\,,\ \ \ \ \ \ \hat{\Y}=-\frac{16M}{r^3}\,.
\end{aligned}\ea
Then, with the similar calculation, one can obtain the CA complexity growth rate,
\ba\begin{aligned}
\dot{I}&=16\W_2M \lf[\frac{1}{L^2}+\frac{1}{\epsilon^2}-\frac{2M}{\epsilon^3}-\frac{1}{2r_1^2}f(r_1)\ln\lf(-\frac{4f(r_1)l_\text{ct}^2}{r_1^2}\rt)\rt]\,.
\end{aligned}\ea
We can see that for this non-perturbation theory, there might also exist some similar divergences for CA complexity growth rate. And this implies that the divergence can also be found in the non-perturbation theory.

\section{Modified CA conjecture}\label{se3}
According to the above calculation as well as the discussion in \cite{Nally:2019rnw}, we can see that these divergences come from the bulk as well as the spacelike surface $S$ which will approach the singularity. Therefore, if we neglect the bulk and this spacelike surface contributions to the full action, our result will be convergent. Hence, with roughly consideration, we can modify the CA duality as
\ba
\math{C}_\text{mod}=\frac{2 d}{(d+1)\p\hbar}\lf(I_\text{joint}+I_\text{null}\rt)
\ea
for a $(d+2)$-dimensional spacetime. Here the coefficient ensures that the late time limit of the complexity growth rate approaches $2M/\p\hbar$. We can see that this new conjecture will give a convergent holographic complexity growth rate. Moreover, we also need to check whether it also gives expected properties such that it can be regarded as a complexity. To show this, next, we first evaluate the growth rate of this modified complexity in a SAdS-type spacetime, and then check its switchback effect of this modified conjecture.

\subsection{Complexity growth}\label{se31}
In this subsection, we consider a $(d+2)$-dimensional SAdS-type black hole in a general higher curvature gravitational theory. Here the line element can be described by
\ba\begin{aligned}\label{dsd}
ds^2=-f(r)dt^2+\frac{dr^2}{f(r)}+d\S_{k,d}^2\,.
\end{aligned}\ea
And the total action is given by
\ba\begin{aligned}\label{action2}
&I=\int_\math{M}d^{d+2}x \sqrt{-g}F(R_{abcd}) +4\sum_{s}\lf(\int_{\math{B}_s}d^{d+1}x\sqrt{|h|}\Y^{ab}K_{ab}\rt)\\
&+(-1)^\l \int_{\math{C}_{\l}}d^dx\sqrt{\s}\hat{\Y}\h_\l-\int_{\math{N}}d\l d^dx\sqrt{\s}\hat{\Q}\ln\lf(l_{\text{ct}}\Q\rt)\,,
\end{aligned}\ea
with
\ba
\y^{abcd}=\frac{\pd F(R_{abcd})}{\pd R_{abcd}}\,.
\ea
And $k=\{-1,0,1\}$ denotes the $d$-dimensional spherical, planar, and hyperbolic geometry, individually. For the joint contributions, with similar calculation as Sec.2, we can further obtain,
\ba\begin{aligned}
\d I_\text{joint}&=\W_{k,d}\d r \lf.\frac{\pd}{\pd r}\lf[\F(r)\ln \lf(-f(r)\rt)\rt]\rt|_{r=r_1}\\
&=-\frac{\W_{k,d}\d t}{2} \lf[\F(r_1)f'(r_1)+\F'(r_1)f(r_1)\ln \lf(-f(r_1)\rt)\rt]\\
\end{aligned}\ea
where we denoted
\ba
\F(r)=r^d\hat{\Y}(r)
\ea
and used the relation \eq{dr}. Then, we turn to the null boundary contributions. If we adopt the affine parameter for the null generator of the null segments, then, the null surface term will vanish. Therefore, we only need consider the counterterm contributions. By the translation symmetry, there are only two null segments contributing to the action growth. According to the line element \eq{dsd}, one can further obtain $\Q=d/r$. For the past right null segment, the counter term can be shown as
\ba\begin{aligned}
I_\text{ct}^{(1)}=-\W_{k,d}\int_{r_1}^{r_\L}dr \F'(r)\ln\lf(\frac{dl_\text{ct}}{r}\rt)\,,
\end{aligned}\ea
Then, we have
\ba
\d I_\text{ct}^{(1)}=-\frac{\W_{k,d}\d t}{2}\F'(r_1)f(r_1)\ln\lf(\frac{dl_\text{ct}}{r_1}\rt)\,.
\ea
By using the fact $\d I_\text{ct}^{(1)}=\d I_\text{ct}^{(2)}$, one can further obtain
\ba\begin{aligned}
\dot{\math{C}}_\text{mod}&=-\frac{d \W_{k,d}}{(d+1)\p\hbar}\lf[\F(r_1)f'(r_1)+\F'(r_1)f(r_1)\ln \lf(\frac{-d^2l_\text{ct}^2f(r_1)}{r_1^2}\rt)\rt]\,.
\end{aligned}\ea
At the late time limit $r_1\to r_h$, we have
\ba\begin{aligned}\label{dcdt}
\dot{\math{C}}_\text{mod}&=-\frac{d \W_{k,d}}{(d+1)\p\hbar}r_h^d\hat{\Y}(r_h)f'(r_h)\\
&=\frac{2 d TS}{(d+1)\p\hbar}
\end{aligned}\ea
where we used the expressions of the entropy and temperature of this black hole
\ba
S=-2\p\W_{k,d}r_h^d\hat{\Y}(r_h)\,,\ \ \ \ \ T=\frac{1}{4\p}f'(r_h)\,.
\ea

We can note that, at the late times, the growth rate of this modified holographic complexity are always proportional to $TS$. As pointed out in \cite{D.Stanford}, this is expected based on quantum circuit model of complexity: the entropy represents the width of the circuit and the temperature is an obvious choice for the local rate at which a particular qubit interacts.
\begin{figure}
\centering
\includegraphics[width=0.8\textwidth]{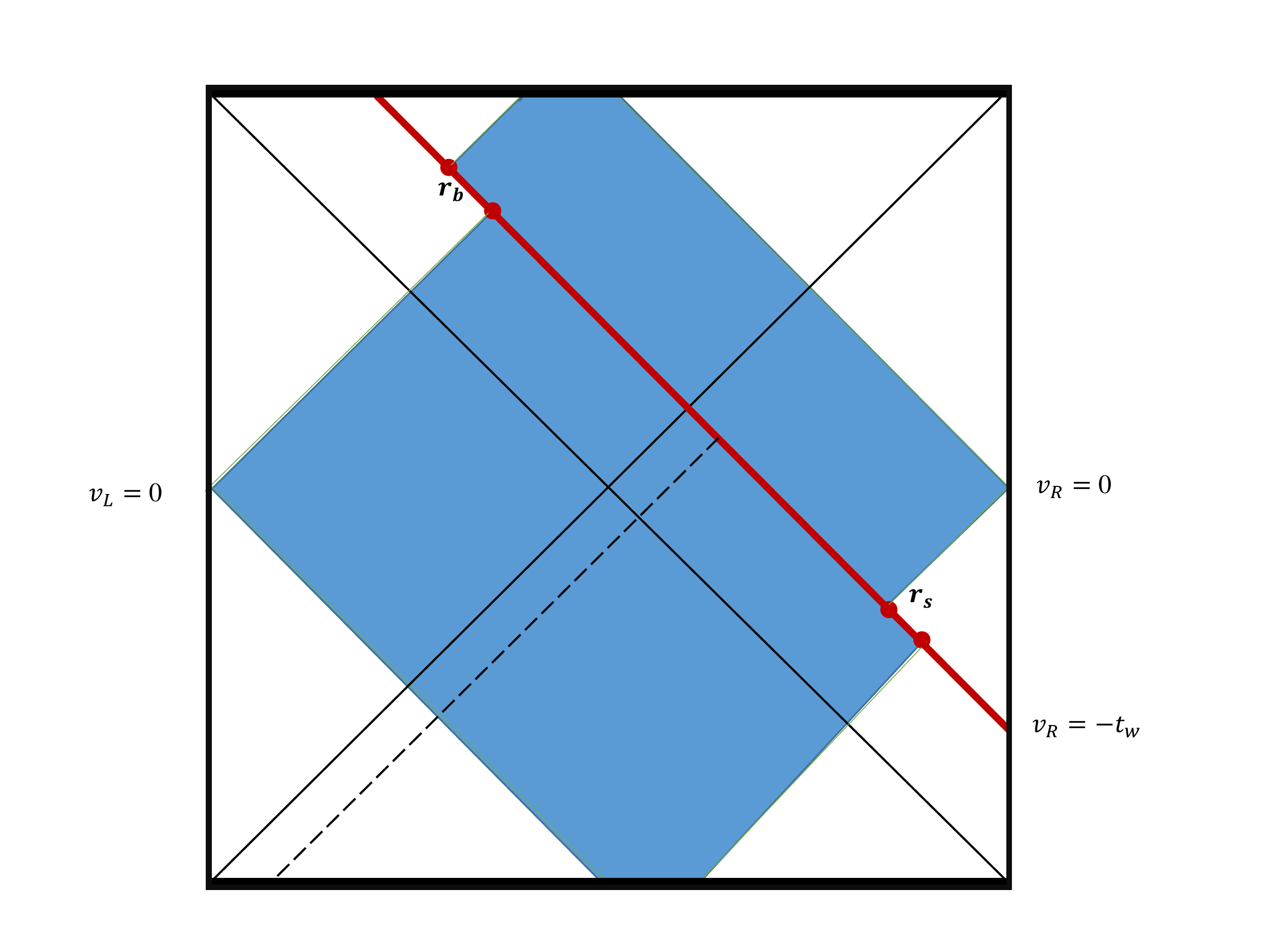}
\caption{ Penrose-like diagram for one shock wave on an SAdS-type black hole geometry.}\label{VaidyaWdW}
\end{figure}

\subsection{Switchback effect}\label{se32}
The switchback effect is an important property for the holographic complexity. It is related the complexity of precursor operators. In this section, to test our conjecture by examining this feature, we consider a Vaidya black hole with thin shell of null fluid collapses in a higher curvature gravitational theory. If we only focus on the light shock wave case, then the black hole's event horizon will shift by a small amount, i.e., we have
\ba
\frac{r_{h,2}}{r_{h,1}}=1+\d\,,
\ea
where the subscripts $1$ and $2$ denote the quantities before and after the shock wave, individually. Then, the scrambling time can be expressed as
\ba
t^*_\text{scr}=\frac{1}{2\p T_1}\ln \frac{1}{\d}\,.
\ea
For the dual quantum system of this black hole, there are two typical regions which are divided by the scrambling time $t=t^*_{\text{scr}}$ after a perturbation is introduced. For $t<t^*_\text{scr}$, the complexity remains essentially unchanged, while for $t>t^*_\text{scr}$, it begins to grow linearly. And our goal in this subsection is to investigate whether this feature can be preserved in our modified CA conjecture.

Following the similar analysis and notation in \cite{Chapman1,Chapman2}, the line element of this shock wave geometry can be written as
\ba\label{dsv}
ds^2=-F(r,v)dv^2+2dr dv +r^2d\S^2_{k,d}\,
\ea
with the blackening factor
\ba\label{Fr}\begin{aligned}
F(r,v)=f_2(r)\math{H}(v+t_w)+f_1(r)\lf[1-\math{H}(v+t_w)\rt]\,,
\end{aligned}\ea
and here we set $r_{h,2}>r_{h,1}$. This line element describes an infinitely thin shell collapse which generates a
shape transition from a SAdS-type black hole with horizon $r=r_{h,1}$ to another one with $r=r_{h,2}$.

For the convenience of later calculations, we would like to introduce the tortoise coordinates as
\ba\begin{aligned}
v_R&<-t_w:\ \ \ \ \ r_1^*(r)=-\int_r^\inf \frac{dr}{f_1(r)}\,,\\
v_R&>-t_w:\ \ \ \ \ r_2^*(r)=-\int_r^\inf \frac{dr}{f_2(r)}\,.
\end{aligned}\ea
We choose this range of integration to make that both expressions satisfy  the boundary condition $\lim_{r\to\inf}r^*_{1,2}(r)\to 0$. Using these coordinates, one can also define an ``outgoing" null coordinate $u$ and auxiliary time coordinate $t$ as
\ba
u_{i}\equiv v-2r^*_i(r),\ \ \ \ \ \ \ t_i\equiv v-r^*_i(r)\,.
\ea
The switchback effect is revealed in the ``complexity of formation" of the boundary thermofield double state (TFD), which is the extra complexity required to prepare the two copies of the fermionic field theory in the TFD state compared to simply preparing each of the copies in the vacuum state, i.e.,
\ba
\D\math{C}=\math{C}\lf(\bra{\text{TFD}}\rt)-\math{C}\lf(\bra{0}_L\otimes\bra{0}_R\rt)
\ea
In the context of AdS/CFT, the holographic calculation is to evaluate the complexity at $t_L=t_R=0$ for the black hole and subtract that for two copies of the AdS vacuum geometry. Therefore, in the following, it is sufficient to restrict our attention to the case $t_L=t_R=0$. In order to show the switchback effect for the modified conjecture, next, we consider the derivative of the complexity of formation with respect to $t_w$ (the slope of the complexity of formation). And since the holographic complexity for the AdS vacuum geometry is time-independent, the slope of complexity of formation can be expressed as
\ba
\frac{d \D\math{C}_\text{mod}}{d t_w}=\frac{d \math{C}_\text{mod}}{d t_w}
\ea
To examine the switchback effect, next we consider two limit regions $t \ll t^*_\text{scr}$ and $t \gg t^*_\text{scr}$.
For the region $t_w\ll t^*_\text{scr}$, by considering the shift symmetry and the light shock wave condition $f_1\simeq f_2$, the complexity of formation will vanish, i.e.,
\ba
\lf.\frac{d\D \math{C}_\text{mod}}{dt_w}\rt|_{t_w\ll t^*_\text{scr}}=0
\ea

For another region $t_w\gg t^*_\text{scr}$, one can verify that all of the light sheets will sent into singularity. And the geometry of the WDW patch is characterized by the dynamical points: $r_s$ and $r_b$, the point where the null shell crosses the past right and future left boundaries. By using the tortoise coordinates, these positions yield
\ba\begin{aligned}
t_w+2r_2^*(r_s)=0\,,\ \ \ \ \ t_w+2r_1^*(r_b)=0\,.
\end{aligned}\ea
Then, the derivative of these dynamical points with respect to $t_w$ is given by
\ba
\frac{d r_s}{d t_w}=-\frac{f_2(r_s)}{2}\,,\ \ \ \ \frac{d r_b}{d t_w}=-\frac{f_1(r_b)}{2}\,.
\ea

Following the standard prescription proposed by \cite{Chapman1,Chapman2}, we shall apply the affine parameter for null generator of null segments. As a consequence, the contributions from the corners at $r_{s/b}$ , as well as all of the null segments will vanish. By considering the symmetries of this spacetime, the time-dependent contributions only comes from the counter terms of the past right and future left null boundaries. We first consider the past right segment of the WDW patch, the relevant null normals to these boundaries can be defined as
\ba
k^{(p)}_a=H(r,v)\lf[-(dv)_a+\frac{2}{F(r,v)}(dr)_a\rt]
\ea
with affine parameters, where we denote
\ba
H(r,v)=\a_2 \math{H}(r-r_s)+\a_1\lf[1-\math{H}\lf(r-r_s\rt)\rt]\,.
\ea
By demanding that the null boundary is affinely parameterized across the shock wave, we have\cite{Chapman1}
\ba
\frac{\a_1}{\a_2}=\frac{f_1(r_s)}{f_2(r_s)}\,.
\ea
Meanwhile, we will fix the parameter in infinity, i.e., we fix $\a_2$ when we change the shock wave $t_w$. Due to $k^a=\lf(\frac{\pd}{\pd \l}\rt)^a$, one can obtain $dr/d\l=H(r,v)$. Then, by using the expressions of $\Q$ and $\hat{\Q}$, one can further obtain
\ba
\Q=\frac{d H(r,v)}{r}\,,\ \ \ \ \hat{\Q}= \hat{\Q}_2\math{H}(r-r_s)+\hat{\Q}_1\lf[1-\math{H}\lf(r-r_s\rt)\rt]\,.
\ea
with
\ba
\hat{\Q}_i=\frac{\a_i}{r^d}\F_i'(r)
\ea

Whence, the counterterm contribution from the past right null boundary can be written as
\ba\label{Sct1}\begin{aligned}
&I_\text{ct}^{(1)}=-\W_{k,d}\int_{0}^{r_\L}d\l r^d\hat{\Q}\ln\left(\frac{dl_{ct}H(r,v)}{r}\right)\\
&=-\W_{k,d}\int_{0}^{r_\L}dr \F'(r)\ln\left(\frac{dl_{ct}H(r,v)}{r}\right)\\
&=-\W_{k,d}\lf[\int_{0}^{r_\L}dr\F'(r)\ln\left(\frac{dl_{ct}\a_2}{r}\right)
+\F_1(r_s)\ln\left(\frac{f_1(r_s)}{f_2(r_s)}\right)\rt]\,.
\end{aligned}
\ea
Then, we have
\ba
\frac{d I_\text{ct}^{(1)}}{d t_w}=\frac{\W_{k,d}}{2}\lf[\F_1'(r_s)f_2(r_s)\ln\lf(\frac{f_1(r_s)}{f_2(r_s)}\rt)
+\F_1(r_s)\lf(\frac{f_2(r_s)f_1'(r_s)}{f_1(r_s)}-f_2'(r_s)\rt)\rt]\,.
\ea
With same calculation, one can further obtain the counterterm contribution from the left future null segment as
\ba
\frac{d I_\text{ct}^{(2)}}{d t_w}=\frac{\W_{k,d}}{2}\lf[\F_2'(r_b)f_1(r_b)\ln\lf(\frac{f_2(r_b)}{f_1(r_b)}\rt)
+\F_2(r_b)\lf(\frac{f_1(r_b)f_2'(r_b)}{f_2(r_b)}-f_1'(r_b)\rt)\rt]
\ea

By summing the various expressions above, the slope of the complexity of formation can be written as
\ba\begin{aligned}
\frac{d\D\math{C}_\text{mod}}{dt_w}&=\frac{d\W_{k,d}}{(d+1)\p\hbar}\lf[\F_1'(r_s)f_2(r_s)\ln\lf(\frac{f_1(r_s)}{f_2(r_s)}\rt)
+\F_2'(r_b)f_1(r_b)\ln\lf(\frac{f_2(r_b)}{f_1(r_b)}\rt)\rt.\\
&\left.+\F_1(r_s)\lf(\frac{f_2(r_s)f_1'(r_s)}{f_1(r_s)}-f_2'(r_s)\rt)
+\F_2(r_b)\lf(\frac{f_1(r_b)f_2'(r_b)}{f_2(r_b)}-f_1'(r_b)\rt)\right]\,.
\end{aligned}\ea
At the large time limit $t_w\to \inf$, we have $r_s\to r_{h,2}, r_b\to r_{h,1}$. Then, we have
\ba\label{latetimedc}
\left.\frac{d\D\math{C}_\text{mod}}{dt_w}\right|_{t_w\gg t^*_\text{scr}}=-\frac{d\W_{k,d}}{(d+1)\p\hbar}\lf[\F_1(r_{h,2})f_2'(r_{h,2})+\F_2(r_{h,1})f_1'(r_{h,1})\right]\,.
\ea
In order to study the switchback effect, next, we consider the light shock wave case, i.e., the null shell only injects a small amount of energy into the system. In this situation, we have $f_1\simeq f_2$. Then, Eq. \eq{latetimedc} becomes
\ba
\left.\frac{d\D\math{C}_\text{mod}}{dt_w}\right|_{t_w\gg t^*_\text{scr}}=\frac{4 d TS}{(d+1)\p\hbar}\,.
\ea
\begin{figure}
\centering
\includegraphics[width=4.5in,height=3in]{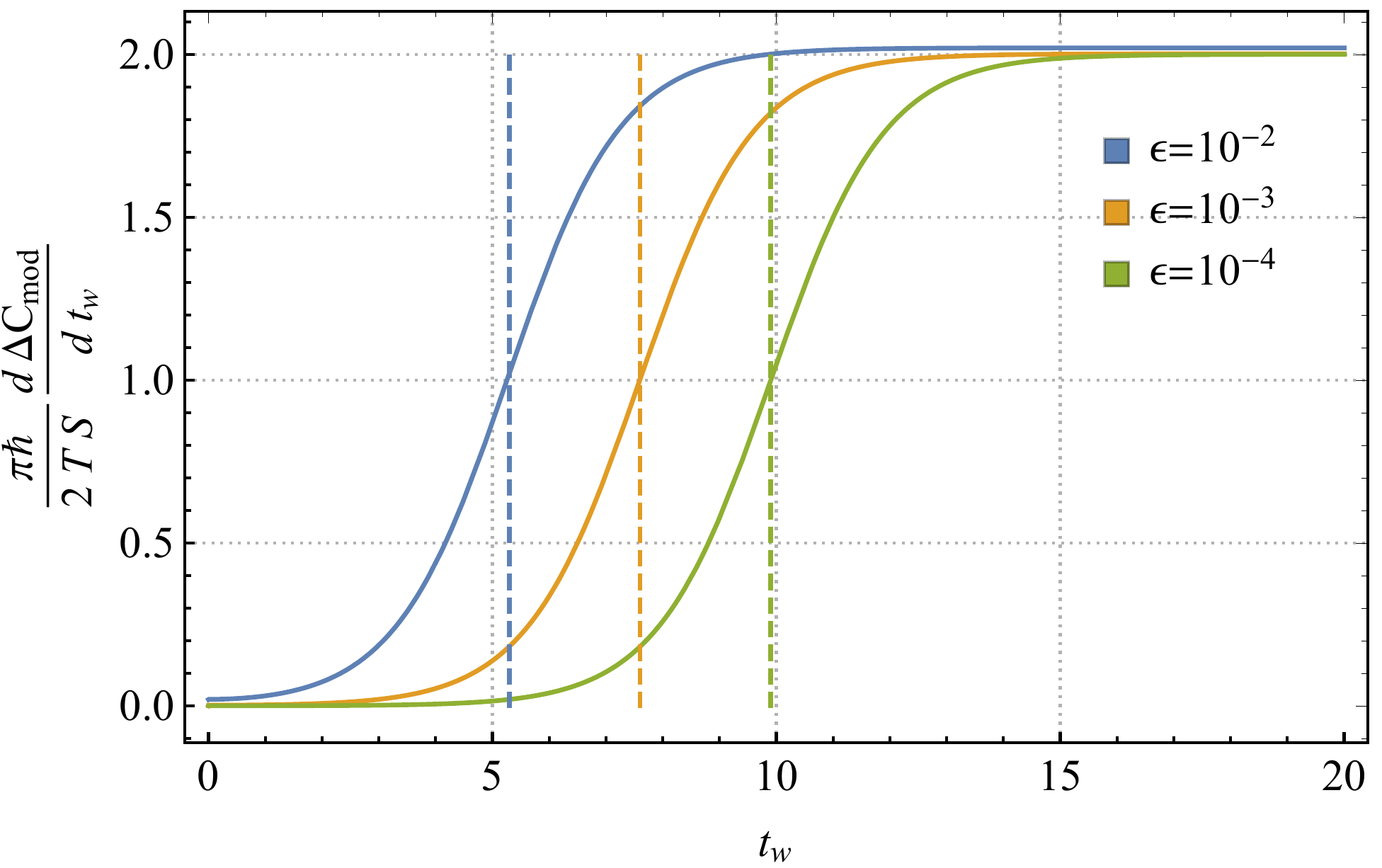}
\caption{The derivative of the complexity of formation with respect to $t_w$ for the BTZ geometry, where we set $L=1,\w_1=1$, and the dishes lines describe the corresponding scrambling time $t=t^*_\text{scr}$.}
\label{switchbackt}
\end{figure}
This result is in agreement with the switchback effect which has been illustrated in the previous literatures \cite{Chapman2,Jiang:2018tlu} for the CA conjecture.

To be specific, in \fig{switchbackt}, we present the slope of the complexity of formation for a light shock wave in BTZ black hole for Einstein gravity. We can see that there exists a scrambling time $t_{\text{scr}}^*$ which is characterized by the energy of the shock wave. And the slope is approximately zero until the $t\simeq t_{\text{scr}}^*$ at which point it rapidly rises to the final constant value. This implies that for the order of the scrambling time $t_{\text{scr}}^*$, the complexity of formation is same as the case of unperturbed state, and in the regime of $t_w>t_{\text{scr}}^*$, it grows linearly with respect to the time $t_w$.
\begin{figure}
\centering
\includegraphics[width=0.8\textwidth]{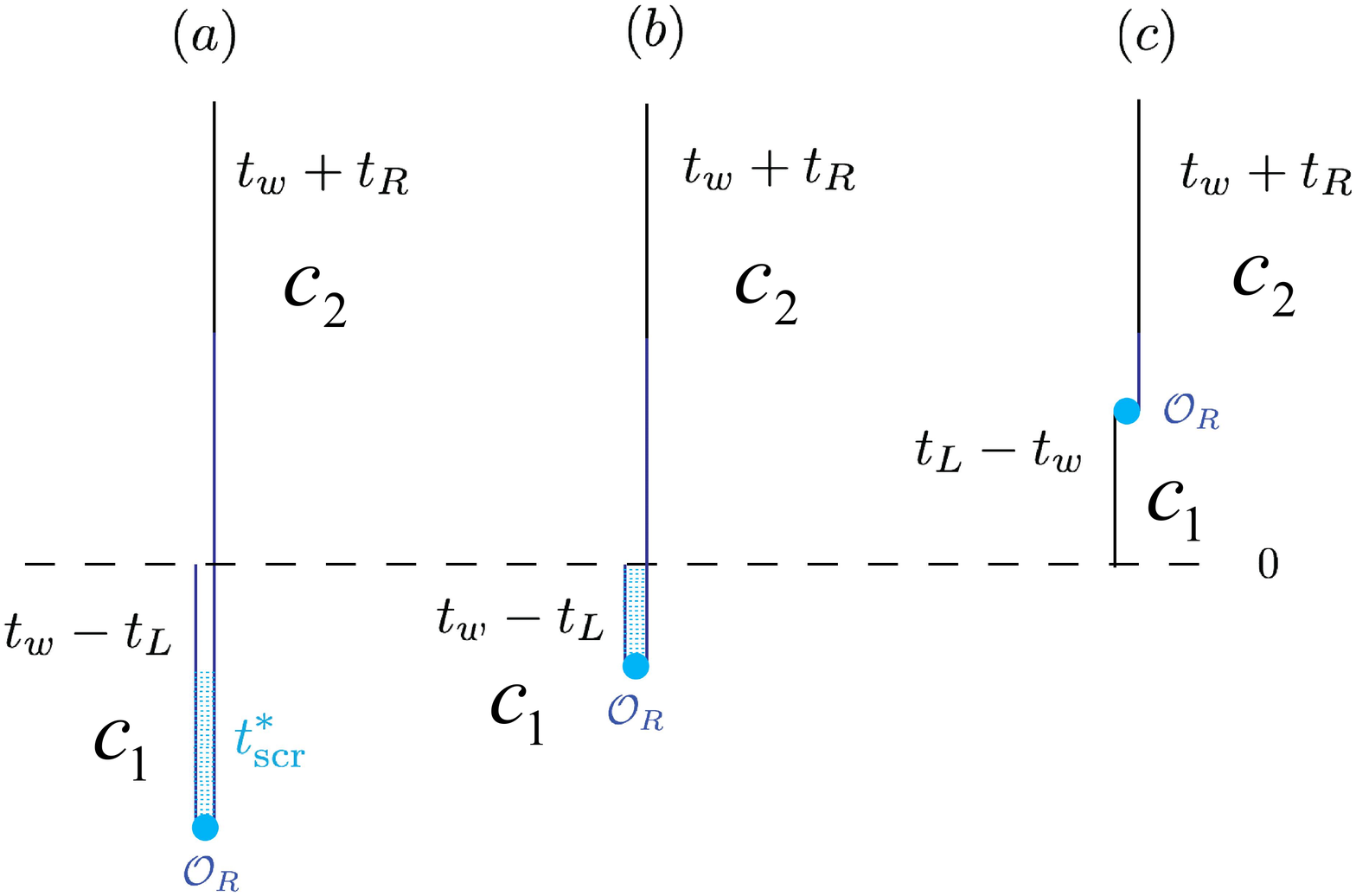}
\caption{A representation of the insertion of a  perturbed operator $\math{O}_R$ at the time $-t_w$ for the TFD state, in analogy to the construction in figure 25 of \cite{Chapman2} as well as figure 6 of \cite{D.Stanford}.}\label{swc}
\end{figure}
\subsection{Circuit analogy}\label{se33}
In this subsection, we would like to investigate the connection between the behaviours of our holographic results and the switchback effect of the circuit model. As discussed in \cite{Chapman2,D.Stanford}, evolving the perturbed state independently in the left and right times yield the expression
\ba\label{SBP}
\bra{TFD(t_L,t_R)}_\text{pert}=U_R(t_R+t_w)\math{O}_RU_R(t_L-t_w)\bra{TFD}\,,\nn
\ea
where the perturbed operator $\math{O}_R$ is a localized simple operator. $U_R(t)\math{O}_R U_R(-t)=I$ with the identity operator $I$ when $t<t^*_\text{scr}$. This feature is connected to the switchback effect\cite{D.Stanford,SZ} and can provide a deeper explanation of our holographic results.

We denote the rate of the complexity to $c_1$ before the operator $\math{O}_R$ is inserted and $c_2$ after it\cite{Chapman2}. Under the limit of light shock, according to \eq{dcdt}, we have
\ba
c_1\approx c_2\approx\frac{2 d TS}{(d+1)\p\hbar}
\ea
First of all, we consider the case $t_w<t^*_\text{scr}$. When $t_L<t_w$, the process in \eq{SBP} can be illustrated in (b) of \fig{swc}. In this situation, the switchback effect produces a cancellation for the process below the dashed line. Therefore, the complexity is given by
\ba\label{cpert1}
\math{C}_\text{pert}\approx \frac{2 d TS t}{(d+1)\p\hbar}\,,
\ea
where we set $t_L=t_R=t/2$. One can note that this complexity is exactly the result of the eternal case where the cancellation is always valid for the process below the dashed line. When $t_L>t_w$, the process can be illustrated by (c) in \fig{swc}. We can see that there is no opportunity for the switchback effect. Hence, the complexity is also the result of the eternal case which can be described by \eq{cpert1}. As a summary, we find that when $t_w<t^*_\text{scr}$, by virtue of the switch back effect, the complexity is same as that of the unperturbed state.

Then, we consider the case $t_w>t^*_\text{scr}$. When $t_L-t_w>-t^*_\text{scr}$, the complexity shares the same result with the case $t_w<t^*_\text{scr}$. When $t_w-t_L>t^*_\text{scr}$, the process can be illustrated by (a) in \fig{swc}. In this case, the two time-evolution operators cancel out only during the scrambling time. Therefore, the complexity can be written as
\ba
\math{C}_\text{pert}\approx \frac{4 d TS}{(d+1)\p\hbar}\,(t_w-t^*_\text{scr})\,.
\ea
This result shows that the growth rate is very close to zero in the region $t<2(t_w-t^*_\text{scr})$.

Next, we consider the complexity of formation. By setting $t=0$ an using the above equations, one can obtain
\ba
\frac{d\D \math{C}_\text{pert}}{d t_w}\approx\frac{4 d TS}{(d+1)\p\hbar}\math{H}(t_w-t^*_\text{scr})\,.
\ea
Again, this formula also matchs the our holographic case as illustrated in the last subsection in which when $t<t^*_\text{scr}$, the rate of the complexity of formation is close to zero, and when $t>t^*_\text{scr}$, it remains constant.

\section{Conclusion}\label{se4}

It has been found that the CA complexity growth rate is divergent when a higher curvature correction is taken into account\cite{Nally:2019rnw}. To better understand the divergences in the CA conjecture, we studied both the perturbation Einsteinian cubic gravity and non-perturbation Einstein-Weyl gravity. And we found that these divergences can also be found in the non-perturbation theory. In order to obtain expected properties of the holographic complexity, we modified the CA conjecture, where we assume that the complexity of the boundary state equals the boundary actions contributed by the null segments as well as the joints of the WDW patch. Then, we can see this modified holographic complexity will be convergent. As a candidate of the boundary complexity, it is necessary to check whether it also gives expected properties as the circuit complexity, such as the late time growth rate and switchback effect. Therefore, in Sec.\ref{se31}, we first calculate the complexity growth in a SAdS-type spacetime for a general higher curvature gravity. And the late time growth rate of this modified holographic complexity is proportional to $TS$, which is quite in agreement with circuit analysis, where the entropy represents the width of the circuit and the temperature is an obvious choice for the local rate. Hence, at this point, our conjecture is suitable for describing the circuit complexity of the boundary state. Finally, we also investigate the switchback effect of this new conjecture in Vaidya geometry. And we found that our result is actually in agreement with the switchback effect which has been illustrated in the previous literatures for the original CA conjecture, although here all of the contributions at late time comes from the counterterms of the null boundary.

\section*{Acknowledgments}
 This research was supported by NSFC Grants No. 11775022 and 11375026.


\begin{thebibliography}{100}
\bibitem{Takayanagi:2016}
S. Ryu and T. Takayanagi, ``Holographic derivation of entanglement entropy from AdS/CFT,"
Phys. Rev. Lett. {\bf96} 181602 (2006).
\bibitem{VanRaamsdonk:2010pw}
  M.~Van Raamsdonk, ``Building up spacetime with quantum entanglement,''  Gen.\ Rel.\ Grav.\  {\bf 42}  2323 (2010).
\bibitem{Maldacena:2013xja}
J.~Maldacena and L.~Susskind, ``Cool horizons for entangled black holes,'' Fortsch.\ Phys.\ {\bf 61} 781 (2013).
\bibitem{Aharony:1999ti}
  O.~Aharony, S.~S.~Gubser, J.~M.~Maldacena, H.~Ooguri and Y.~Oz, ``Large N field theories, string theory and gravity,''   Phys.\ Rept.\  {\bf 323} 183 (2000).
\bibitem{Einstein:1935}
A. Einstein and N. Rosen, ``The particle problem in the general theory of relativity,'' Phys. Rev. {\bf48} 7377 (1935).
\bibitem{Susskind:2014moa}
  L.~Susskind, ``Entanglement is not enough,''  Fortsch.\ Phys.\  {\bf 64} 49 (2016).
\bibitem{L.Susskind} 
L. Susskind,``Computational complexity and black hole horizons,'' Fortsch. Phys. {\bf64} 24(2016).
\bibitem{D.Stanford} D. Stanford and L. Susskind, ``Complexity and shock wave geometries,'' Phys. Rev. D {\bf90}, 126007(2014).
\bibitem{BrL} A. R. Brown, D. A. Roberts, L. Susskind, B. Swingle and Y. Zhao,``Holographic complexity equals bulk action?'' Phys. Rev. Lett. {\bf116} 191301(2016).
\bibitem{BrD} A. R. Brown, D. A. Roberts, L. Susskind, B. Swingle and Y. Zhao,``Complexity, action, and black holes,'' Phys. Rev. D {\bf93} 086006(2016).
\bibitem{A1}
 J.~Jiang, ``Action growth rate for a higher curvature gravitational theory,'' Phys.\ Rev.\ D {\bf 98}, 086018 (2018).
\bibitem{A2}
 R. G. Cai, S. M. Ruan, S. J. Wang, R. Q. Yang and R. H. Peng, JHEP {\bf1609} (2016).
\bibitem{A3}
L. Lehner, R. C. Myers, E. Poisson and R. D. Sorkin, ``Gravitational action with null boundaries'' Phys. Rev. D {\bf94}, 084046(2016).
\bibitem{A4}
D. Carmi, S. Chapman, H. Marrochio, R. C. Myers and S. Sugishita, JHEP {\bf1711} 188(2017).
\bibitem{Fan:2019mbp}
  Z.~Y.~Fan and M.~Guo, ``Holographic complexity and thermodynamics of AdS black holes,'' arXiv:1903.04127.
\bibitem{A51}
Z.~Y.~Fan and M.~Guo, ``Holographic complexity under a global quantum quench,'' arXiv:1811.01473.
\bibitem{A5}
Z.~Y.~Fan and M.~Guo, ``On the Noether charge and the gravity duals of quantum complexity,'' JHEP {\bf 1808}, 031 (2018).
\bibitem{A6}
Y.~S.~An, R.~G.~Cai and Y.~Peng, ``Time Dependence of Holographic Complexity in Gauss-Bonnet Gravity,'' Phys.\ Rev.\ D {\bf 98}, 106013 (2018).
\bibitem{A7}
Y.~S.~An and R.~H.~Peng, ``Effect of the dilaton on holographic complexity growth,'' Phys.\ Rev.\ D {\bf 97} 066022 (2018).
\bibitem{A8}
A. Reynolds and S. F. Ross, Class. ``Complexity in de Sitter Space,'' Quant. Grav. {\bf34},175013(2017).
\bibitem{A9}
S. Chapman, H. Marrochio and R. C. Myers, ``Complexity of Formation in Holography,'' JHEP {\bf1701} 062(2017).
\bibitem{A38}
X.~H.~Feng and H.~S.~Liu, ``Holographic Complexity Growth Rate in Horndeski Theory,'' arXiv:1811.03303.
\bibitem{A10}
D. Carmi, R. C. Myers and P. Rath,``Comments on Holographic Complexity,'' JHEP {\bf1703} 118(2017).
\bibitem{A11}
M. Alishahiha, ``Holographic Complexity,''  Phys. Rev. D {\bf92}  126009 (2015).
\bibitem{A12}
C. A. Agon, M. Headrick and B. Swingle,``Subsystem Complexity and Holography,'' arXiv:1804.01561.
\bibitem{A13}
O. Ben-Ami and D. Carmi, ``On Volumes of Subregions in Holography and Complexity,'' JHEP {\bf1611}, 129 (2016).
\bibitem{A14}
Y.~Zhao, ``Uncomplexity and Black Hole Geometry,'' Phys.\ Rev.\ D {\bf 97}, 126007 (2018)
\bibitem{A15}
Z. Fu, A. Maloney, D. Marolf, H. Maxfield and Z. Wang, ``Holographic complexity is nonlocal,'' JHEP {\bf02} 072(2018).
\bibitem{A16}
M.~Alishahiha, A.~Faraji Astaneh, M.~R.~Mohammadi Mozaffar and A.~Mollabashi, ``Complexity Growth with Lifshitz Scaling and Hyperscaling Violation,'' JHEP {\bf 1807} (2018) 042
\bibitem{A17}
J. Couch, S. Eccles, W. Fischler and M. L. Xiao,``Holographic complexity and noncommutative gauge theory,'' JHEP {\bf1803} 108(2018).
\bibitem{A18}
  B.~Swingle and Y.~Wang, ``Holographic Complexity of Einstein-Maxwell-Dilaton Gravity,''  JHEP {\bf 1809} 106 (2018).
\bibitem{A19}
M. Moosa, ``Evolution of Complexity Following a Global Quench,'' JHEP {\bf1803} 031(2018).
\bibitem{A20}
B.~Chen, W.~M.~Li, R.~Q.~Yang, C.~Y.~Zhang and S.~J.~Zhang, ``Holographic subregion complexity under a thermal quench,''
JHEP {\bf 1807} 034 (2018).
\bibitem{A21}
S. Chapman, H. Marrochio and R. C. Myers, JHEP ``Holographic complexity in Vaidya spacetimes. Part I,'' {\bf 1806} 046(2018).
\bibitem{A22}
S.~Chapman, H.~Marrochio and R.~C.~Myers, ``Holographic complexity in Vaidya spacetimes. Part II,'' JHEP {\bf 1806} 114 (2018).
\bibitem{A37}
K. Hashimoto, N. Iizuka, and S. Sugishita, ``Time Evolution of Complexity in Abelian
Gauge Theories - And Playing Quantum Othello Game -'', arXiv:1707.03840.
\bibitem{A23}
R. A. Jefferson and R. C. Myers, ``Circuit complexity in quantum field theory,'' JHEP 10
(2017) 107.
\bibitem{A24}
S.~Chapman, M.~P.~Heller, H.~Marrochio and F.~Pastawski,`` Towards Complexity for Quantum Field Theory States''  Phys.\ Rev.\ Lett.\  {\bf 120}, 121602 (2018).
\bibitem{A25}
R.-Q. Yang, ``A Complexity for Quantum Field Theory States and Application in
Thermofield Double States,'' arXiv:1709.00921.
\bibitem{A26}
R. Q. Yang, C. Niu, C. Y. Zhang, and K.-Y. Kim, ``Comparison of holographic and field
theoretic complexities for time dependent thermofield double states,'' JHEP {\bf02} 082 (2018).
\bibitem{A27}
 R.~Q.~Yang, Y.~S.~An, C.~Niu, C.~Y.~Zhang and K.~Y.~Kim,``More on complexity of operators in quantum field theory,'' arXiv:1809.06678.
\bibitem{A28}
A. R. Brown and L. Susskind, ``Second law of quantum complexity'' Phys. Rev.  D {\bf 97} 086015 (2018).
\bibitem{A39}
A. P. Reynolds and S. F. Ross, ``Complexity of the AdS Soliton,'' arXiv:1712.03732.
\bibitem{A29}
P. Caputa, N. Kundu, M. Miyaji, T. Takayanagi and K. Watanabe,``Liouville Action as Path-Integral Complexity: From Continuous Tensor Networks to AdS/CFT,'' JHEP {\bf 1711} 097 (2017).
\bibitem{A30}
R. Khan, C. Krishnan, and S. Sharma, ``Circuit Complexity in Fermionic Field Theory,''
arXiv:1801.07620.
\bibitem{A31}
M.~Guo, J.~Hernandez, R.~C.~Myers and S.~M.~Ruan, ``Circuit Complexity for Coherent States,''  JHEP {\bf 1810} 011 (2018).
\bibitem{A32}
J.~Jiang, J.~Shan and J.~Yang, ``Circuit complexity for free Fermion with a mass quench,''  arXiv:1810.00537 [hep-th].
\bibitem{A33}
R.~Q.~Yang, Y.~S.~An, C.~Niu, C.~Y.~Zhang and K.~Y.~Kim, ``Principles and symmetries of complexity in quantum field theory,'' arXiv:1803.01797.
\bibitem{A34}
S.~Chapman, J.~Eisert, L.~Hackl, M.~P.~Heller, R.~Jefferson, H.~Marrochio and R.~C.~Myers, ``Complexity and entanglement for thermofield double states,'' arXiv:1810.05151.
\bibitem{A35}
J.~Jiang and X.~Liu, ``Circuit Complexity for Fermionic Thermofield Double states,'' Phys.\ Rev.\ D {\bf 99}, 026011 (2019)
\bibitem{A36}
L.~Hackl and R.~C.~Myers, ``Circuit complexity for free fermions,'' JHEP {\bf 1807} 139 (2018).
\bibitem{Lloyd} S. Lloyd,``Ultimate physical limits to computation,'' Nature {\bf406} 1047(2000).
\bibitem{Nally:2019rnw}
  R.~Nally,``Stringy Effects and the Role of the Singularity in Holographic Complexity,'' arXiv:1902.09545
\bibitem{BC} 
P. Bueno and P.A. Cano,``Einsteinian cubic gravity,'' Phys. Rev. D {\bf94}, 104005(2016).
\bibitem{Jiang3}
J.~Jiang and H.~Zhang,``Surface term, corner term, and action growth in F(Riemann) gravity theory,'' arXiv:1806.10312.
\bibitem{Mann}
R. A. Hennigar and R. B. Mann,``Black holes in Einsteinian cubic gravity,'' Phys. Rev. D {\bf 95}, 064055(2017)
\bibitem{Chapman1}
S. Chapman, H. Marrochio and R. C. Myers, JHEP ``Holographic complexity in Vaidya spacetimes. Part I,'' {\bf 1806} 046(2018).
\bibitem{Chapman2}
S.~Chapman, H.~Marrochio and R.~C.~Myers, ``Holographic complexity in Vaidya spacetimes. Part II,'' JHEP {\bf 1806} 114 (2018).
\bibitem{Jiang:2018tlu}
  J.~Jiang, ``Holographic complexity in charged Vaidya black hole,''  Eur.\ Phys.\ J.\ C {\bf 79} 130 (2019)
\bibitem{SZ} L. Susskind and Y. Zhao, ``Switchbacks and the Bridge to Nowhere,'' arXiv:1408.2823.
\end{thebibliography}
\end{document}